\documentclass[global,twocolumn,final]{svjour}
\usepackage[latin1]{inputenc}
\usepackage{graphicx}

\usepackage{subfigure}
\usepackage[pdfusetitle,bookmarksnumbered,plainpages=false]{}

\journalname{DOI:10.1007/s00340-011-4563-7, Final publication available at springerlink.com \\}
\begin{document}
\title{All-optical ion generation for ion trap loading}
\author{Kevin Sheridan\inst{1} \and Wolfgang Lange\inst{1} \and Matthias Keller\inst{1}% etc
}                     
\institute{Department of Physics and Astronomy, University of Sussex, Falmer, BN1 9QH, England.}
\date{Received: date / Revised version: date}
\maketitle

%%%ABSTRACT%%%
%%%%%%%%%%%%%%%%%%%%%%%%%%%%%%%%%%%%%%%%%%%%%%%%%%%%%%%%%%%%%%%%%%%%%%%%%%%%%%%
\begin{abstract}
We have investigated the all-optical generation of ions by photo-ionisation of atoms generated by pulsed laser ablation. A direct comparison between a resistively heated oven source and pulsed laser ablation is reported. Pulsed laser ablation with 10~ns Nd:YAG laser pulses is shown to produce large calcium flux, corresponding to atomic beams produced with oven temperatures greater than 650~K. For an equivalent atomic flux, pulsed laser ablation is shown to produce a thermal load more than one order of magnitude smaller than the oven source. The atomic beam distributions obey Maxwell-Boltzmann statistics with most probable speeds corresponding to temperatures greater than 2200~K. Below a threshold pulse fluence between 280~mJ/cm$^2$ and 330 mJ/cm$^2$, the atomic beam is composed exclusively of ground state atoms. For higher fluences ions and excited atoms are generated. 
\end{abstract}

%%%INTRODUCTION%%%
%%%%%%%%%%%%%%%%%%%%%%%%%%%%%%%%%%%%%%%%%%%%%%%%%%%%%%%%%%%%%%%%%%%%%%%%%%%%%%
\section{Introduction}
\label{sec:intro}

The manipulation of trapped ions has become an important experimental technique for quantum information processing \cite{Leibfried,Haffner,Leibfried2}, high resolution spectroscopy \cite{Yb+,Hg+,Al+} and quantum simulation \cite{Tobias,Roos}. Single trapped ions have been shown to be viable qubits in quantum logic gates, the building blocks of a quantum computer \cite{Steane}. In recent years, the scaling-up of ion traps has been proposed through the creation of microtrap arrays \cite{Kielpinski,Stick,Brownnutt}. However, as ion traps become smaller the experimental difficulties presented by the loading process become more critical. Also, combining ion traps with ultra high quality optical cavities \cite{Keller1,Keller2,Russo} poses severe requirements on the loading process as any contamination of the mirror surfaces will deteriorate the cavity quality.

Commonly, ion traps are loaded using a neutral atomic beam created through effusion from a resistively heated oven and ionised either by electron impact or photo-ionisation in the trap centre. Ion trapping experiments typically aim to load single ions or small numbers of ions and thus a large continuous flux of atoms passing through the trap structure is undesirable. Atoms from the beam may become deposited on the trap electrodes or dielectric surfaces in the set-up, leading to local patch potentials. The build-up of contaminants on the trap electrodes has been shown to significantly increase the heating rate of trapped ions over time \cite{DeVoe,HaffnerT}. This is particularly pertinent for experiments seeking to cool trapped ions to the motional ground state. 

The loading efficiency and isotope selectivity that resonant two-step photo-ionisation offers has made it the preferred technique over electron impact ionisation for the clean loading of ion traps. Loading efficiencies up to five orders of magnitude larger \cite{Lucas,Drewsen} means that a smaller flux of neutral atoms is required, resulting in a decrease in the contamination of the trap electrodes \cite{Gulde}.

The atomic flux produced by effusive ovens is continuous and cannot be quickly switched off, thus inevitably producing an excess of evaporated atoms. In addition, the presence of a hot oven at typically several hundred degrees Celcius in the close confines of a micro-scale ion trap array may cause damage. The production of atomic beams through pulsed laser ablation (PLA) eliminates the need for an oven in the experimental design. The ablation of metal surfaces has been studied extensively (see for example \cite{Ashfold,Haglund}) and with the advances of high power pulsed diode lasers the method may become suitable for complete integration in ion chips.

Highly efficient ion trap loading has been demonstrated for calcium ablated with Nd:YAG laser pulses in a large trap \cite{Hendricks} and for the loading of strontium into a surface trap with a frequency trippled Nd:YAG laser \cite{Leibrandt}. Employing high energy UV laser pulses resulted in the generation of ions and excited atoms in the ablation process which makes this method less suitable for isotope selective photoionisation. In contrast, Hendricks et al. demonstrated that thermal ablation with a Nd:YAG laser can produce atoms entirely in their ground state if the fluence is below a threshold around 300~mJ/cm$^2$.

PLA utilises high energy laser pulses which strike a target material and heat a strongly localised region to high temperatures. This creates discrete bunches of atoms leaving the target surface in response to each laser pulse. Together with resonant photo-ionisation, ions can be produced isotope selectively.

Combining PLA with resonant photo-ionisation provides unparalleled control over the ion trap loading process. The evaporation of atoms through the ablation may be maintained as long as required to produce the desired number of trapped ions. Thus, after detecting the correct number of ions the ablation process can be stopped instantaneously. By contrast, evaporation during the heat up and cool down phase of an oven leads inevitably to an excess of released atoms. Due to the high temperatures of the ablated atoms, the loading may be less efficient compared to a conventional oven for shallow traps. However, D.R. Leibrandt et al. \cite{Leibrandt} have demonstrated the loading of a surface trap with a trap depth as low as 40~meV.

In this paper we present a detailed experimental investigation of this all optical ion generation technique and we present an analysis of the suitablity of PLA for ion trap loading.
We directly compare the flux and velocity distributions of atomic beams created by PLA and an effusive oven. By analysing the distribution of the kinetic energy, we evaluate the efficiency of the loading process.

%%%EXPERIMENT%%%
%%%%%%%%%%%%%%%%%%%%%%%%%%%%%%%%%%%%%%%%%%%%%%%%%%%%%%%%%%%%%%%%%%%%%%%%%%%%%%%% 
\section{Experiment}
\label{sec:experiment}

\begin{figure}
\begin{center}
\resizebox{0.50\textwidth}{!}{\includegraphics{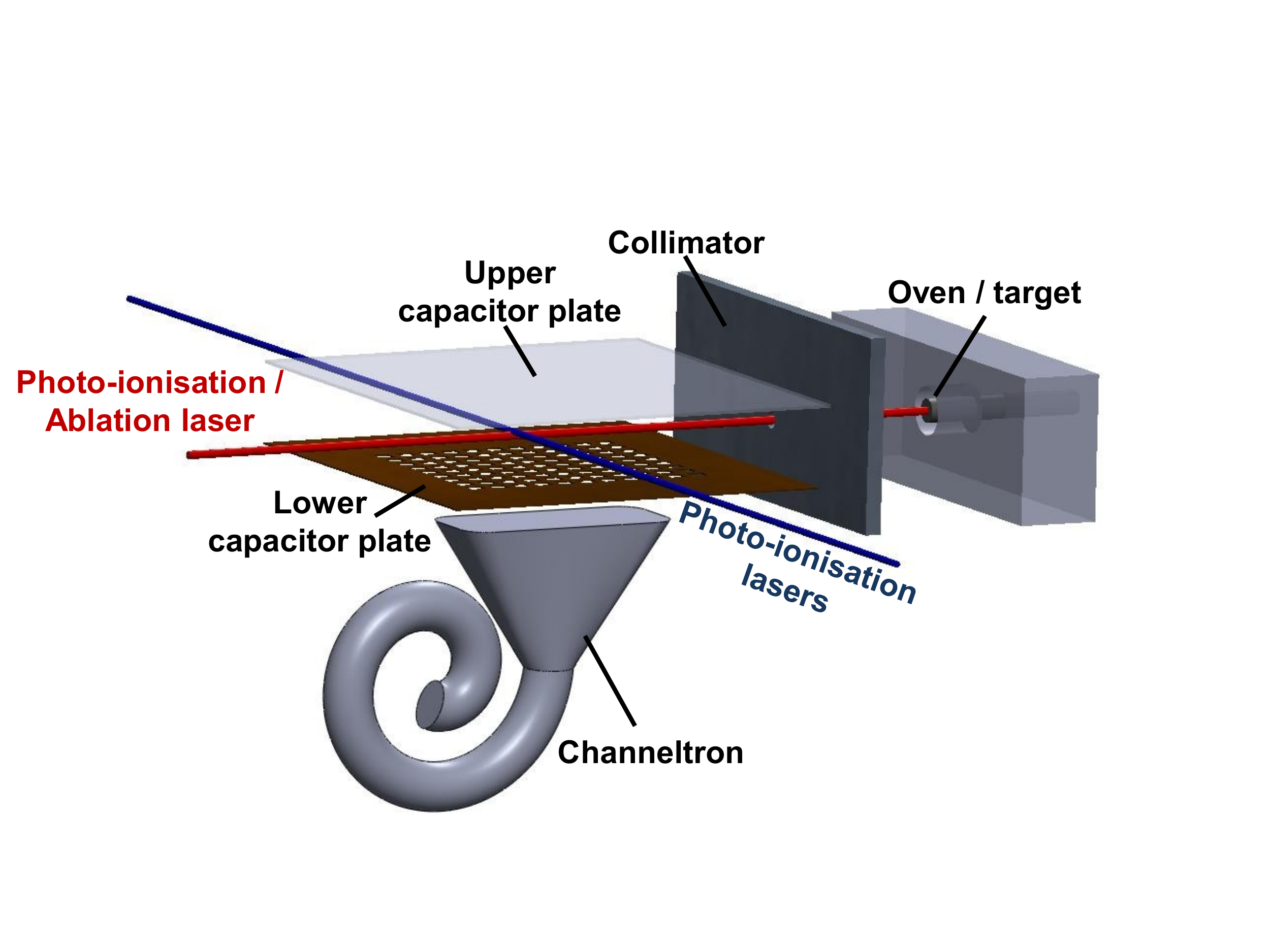}}
\end{center}
\caption{Experimental set-up within the vacuum chamber. The atomic flux originating at the calcium source, either the resistively heated oven or the calcium target, is collimated into an atomic beam by the 250~$\mu$m aperture. The 1064~nm ablation laser beam is directed along the atomic beam axis and normal to the calcium target surface. The photo-ionisation lasers may be aligned orthogonal or counter-propagating (when the ablation laser beam is not in use) or both with respect to the atomic beam axis. Calcium atoms are photo-ionised in the region between the capacitor plates and detected at the channeltron.}
\label{fig:exp_setup}      
\end{figure}

In order to compare the atomic beams created by a resistively heated oven and PLA, the set-up is designed to reproducibly exchange an oven and an ablation target. A schematic of the experimental set-up in use is shown in Fig. \ref{fig:exp_setup}. The sources can be exchanged without altering any other component of the experimental set-up.

Calcium atoms emitted from the source are collimated by a 250~$\mu$m aperture to achieve a beam divergence of smaller than 3$^{\circ}$. The calcium atoms passing through the aperture are photo-ionised 20~mm behind the collimator, above a channeltron ion detector, to measure the flux. The channeltron is mounted below a grounded metal mesh in order to shield the ions from the high voltage which is needed for the operation of the ion detector. The field from a positively charged plane metal plate above the ionisation point accelerates the ionized atoms towards the channeltron.
In order to photo-ionise the neutral atoms, a 423~nm laser excites calcium atoms in the beam from the 4s$\:^1$S$_0$ ground state to the 4p$\:^1$P$_1$ excited state. This is followed by ionisation by a 375~nm laser which ionises the excited atoms \cite{Lucas,Gulde}. 

Both laser beams are overlapped and collimated with a power of 500~$\mu$W for the 423~nm beam and 1~mW for the 375~nm beam. Both laser beams have a measured waist of 250~$\pm$ 15~$\mu$m. One set of photo-ionisation lasers is aligned orthogonal to the atomic beam to minimise Doppler broadening. Another set of photo-ionisation lasers is aligned anti-parallel to the atomic beam by optimising the transmission through the collimator and the oven aperture.

The standard way of creating an atomic beam is effusion from a resistively heated oven. The oven used in this experiment is formed by a 13~mm long tantalum tube with a diameter of 1~mm. The tube is packed with calcium metal filings, crimped shut at one end and left open at the other. A tantalum wire is spot welded to the tube and a current is passed through the wire, resistively heating the oven. Typical oven currents of 1-2~A correspond to an electric power of about 1-4~W.

Alternatively, the atomic beam can be created by PLA of a solid calcium target.
The ablation laser used in this experiment is a Q-switched 1064~nm Nd:YAG laser (Newport BL-10-106Q). The laser can operate with repetition rates between 1~Hz and 160~kHz, has an M$^2$ value less than 1.2 with an average pulse width of 10~ns and a maximum pulse energy of 120~$\mu$J. The ablation laser is focused with a single lens onto a calcium target passing through the centre of the collimation aperture. The target is a solid cylinder of calcium metal mounted inside a stainless steel block. It is electrically isolated, so that a potential difference between the target surface and the collimation aperture plate can be applied. This electric field can be used to prevent charged particles created through the ablation process from passing through the aperture, thus allowing us to analyse the composition of the atomic beam. With a measured beam waist of 40~$\pm$ 2~$\mu$m and for a repetition rate of 15~kHz, laser fluences up to 2000~mJ/cm$^2$ can be achieved. The direction of the laser beam can be adjusted by a mirror which is controlled by a 2-dimensional PZT tilting stage to allow for the scanning of the beam position on the target. Following the procedure described in \cite{Hendricks}, small position changes of the ablation laser beam on the target produce a consistent flux of calcium over the time scales used in this experiment.
The channeltron output signal is converted into TTL pulses and recorded by a computer. A photo-ionisation spectrum can thus be measured by scanning the 423~nm laser wavelength and simultaneously measuring the ion detector count rate. Additionally, the time-of-flight of the atoms from the source to the detection region at a distance of 25~$\pm$~2~mm can be measured by recording the delay between the ablation laser trigger pulses and the ion count signal with a digital multiscaler card with nanosecond resolution.

%%%RESULTS%%%
%%%%%%%%%%%%%%%%%%%%%%%%%%%%%%%%%%%%%%%%%%%%%%%%%%%%%%%%%%%%%%%%%%%%%%%%%%%%%%%% 
\section{Results}
\label{sec:results}

\subsection{Resistively Heated Oven}
\label{subsec:results_oven}

\begin{figure}
\begin{center}
\resizebox{0.50\textwidth}{!}{\includegraphics{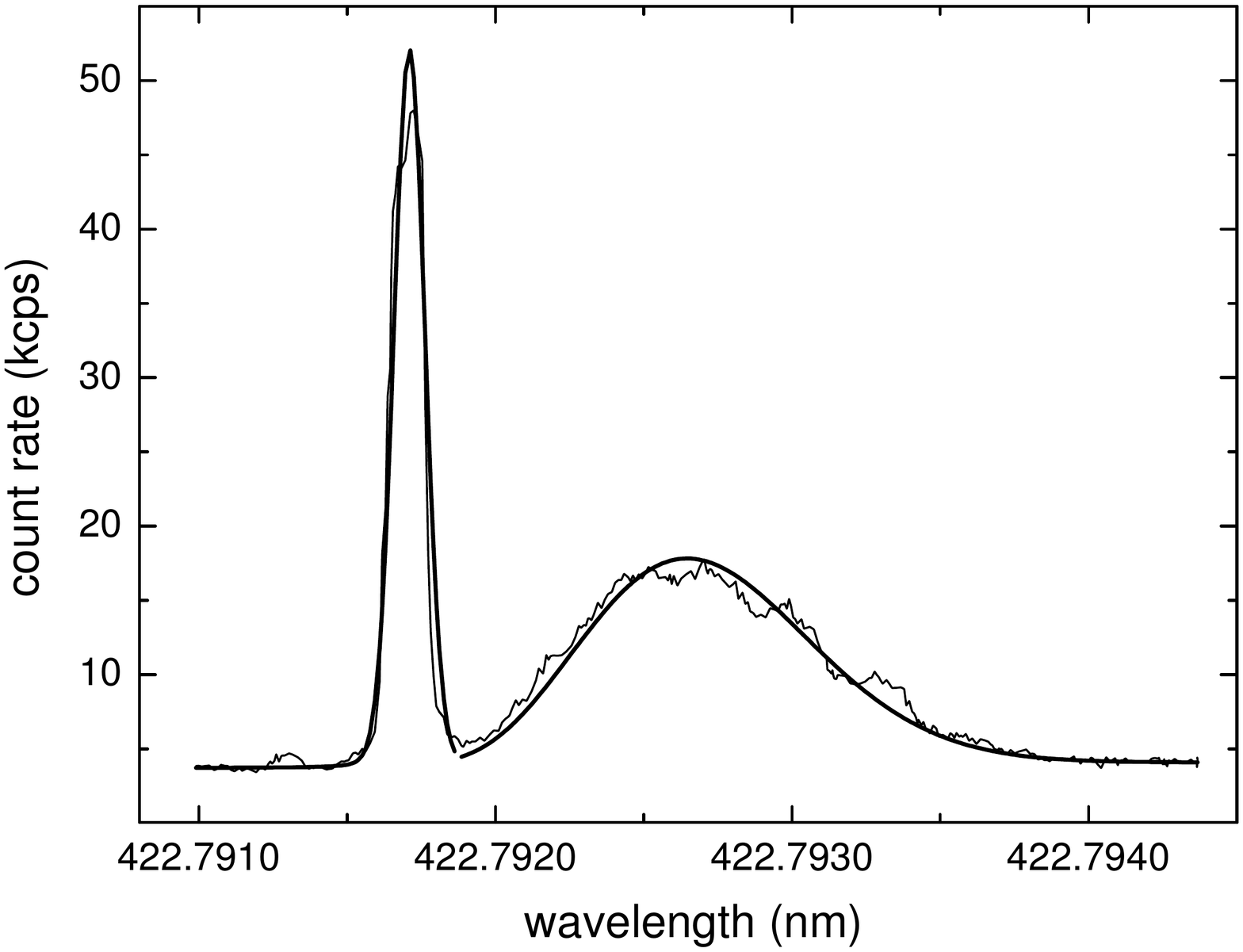}}
\end{center}
\caption{Ion count rate as a function of 423~nm laser wavelength for an oven current of 1.9~A. The Doppler-free absorption profile is obtained by orthogonal photo-ionisation beams while the maximally Doppler broadened profile is obtained by counter-propagating photo-ionisation laser beams. The temperature of the oven is determined by fitting Eq. (\ref{equ:beam_distribution}) to the Doppler broadened absorption profile relative to the measured resonance of the$\:^1$S$_0$$\:\rightarrow$$\:^1$P$_1$ transition.}
\label{fig:oven_spectrum}      
\end{figure}

\begin{figure}
\begin{center}
\resizebox{0.50\textwidth}{!}{\includegraphics{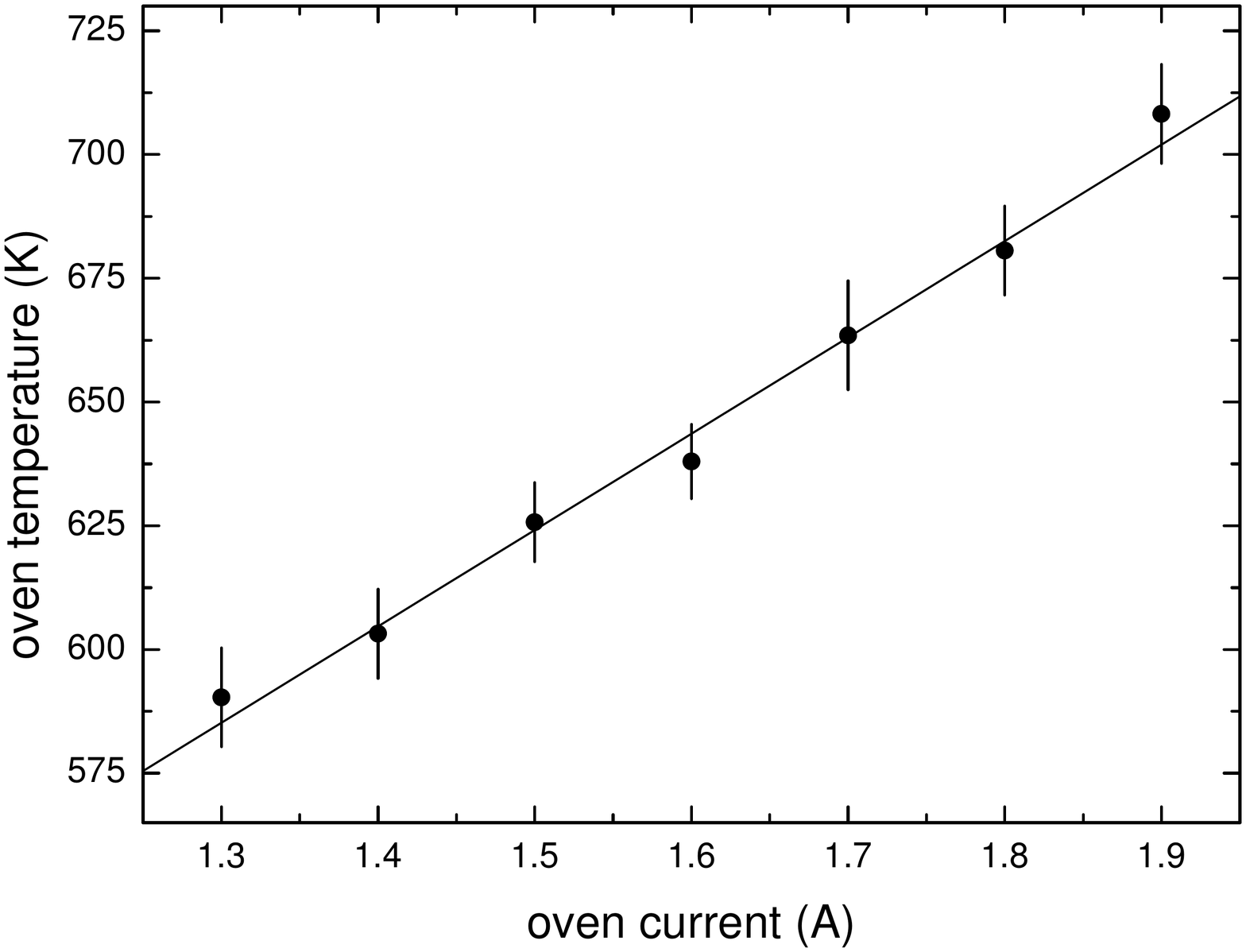}}
\end{center}
\caption{Oven temperature as a function of current. The relationship is linear for the range of currents used in this experiment.}
\label{fig:oven_temp_vs_current}      
\end{figure}

\begin{figure}
\begin{center}
\resizebox{0.50\textwidth}{!}{\includegraphics{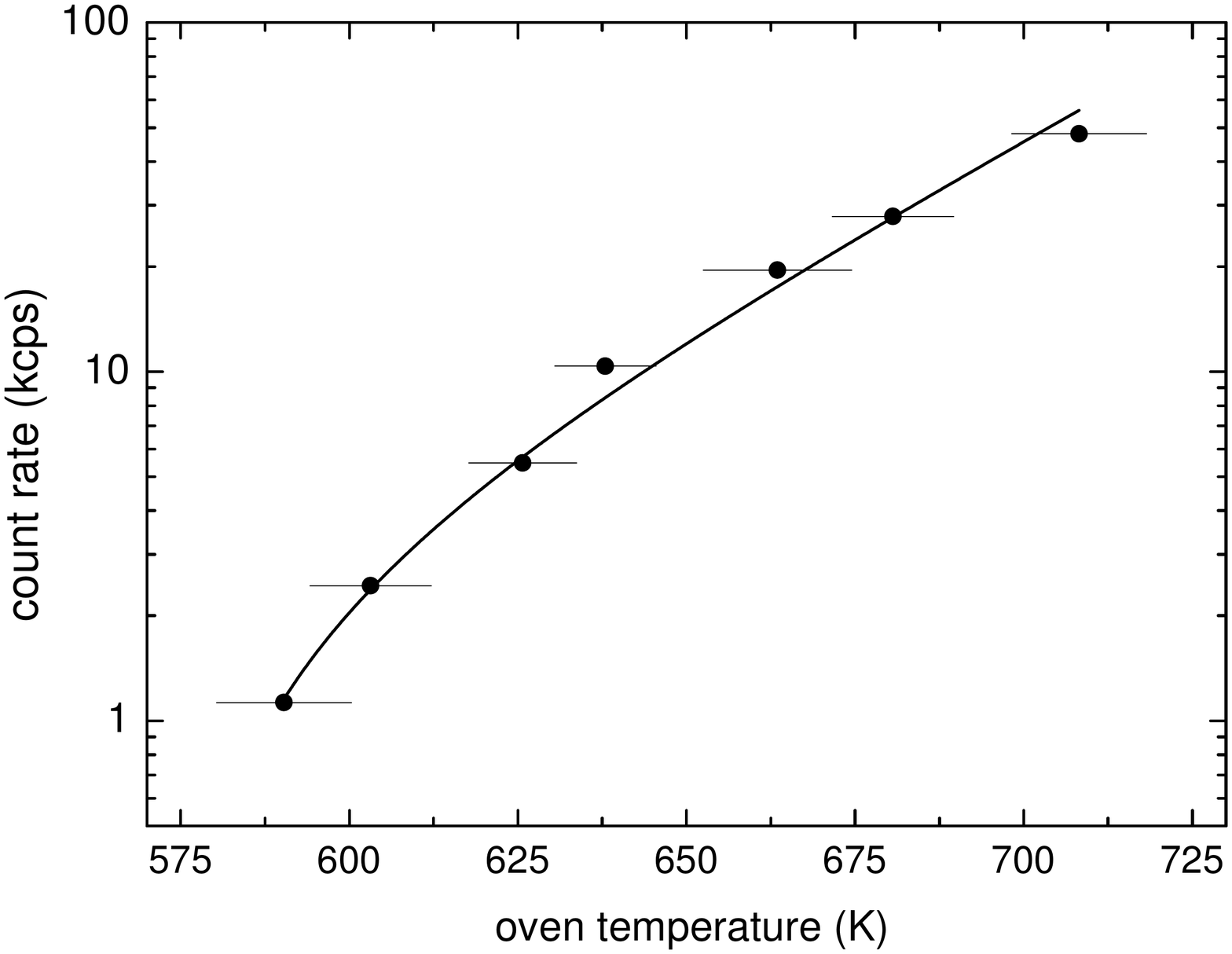}}
\end{center}
\caption{Ion count rate as a function of oven temperature. Oven temperatures are determined from the Doppler broadened absorption profile and count rate is determined by the amplitude of the minimally broadened$\:^1$S$_0$$\:\rightarrow$$\:^1$P$_1$ transition peak. A function has been fit to the data according to the model presented in \cite{Pilling}.}
\label{fig:oven_flux}     
\end{figure}

In order to obtain a measure of the atomic flux effusing from the oven and to measure the wavelength of the$\:^1$S$_0$$\:\rightarrow$$\:^1$P$_1$ transition, the orthogonal and counter-propagating photo-ionisation lasers are scanned simultaneously over a 5~GHz range. Fig. \ref{fig:oven_spectrum} shows the photo-ionisation spectrum for an oven current of 1.9~A. The sharp peak at low wavelength is centered at the transition wavelength and originates from the PI of atoms by the orthogonal laser as there is no first order Doppler shift. 
A Voigt profile is fit to the resonance peak to obtain the width and position. Due to the large laser powers, the width is mainly determined by the saturation broadening of the transition. Doppler broadening of the resonance due to the divergence of the atomic beam is negligible.
The broad peak at longer wavelengths in Fig. \ref{fig:oven_spectrum} is a result of the interaction of the atoms with the counter-propagating laser.
The spectrum represents the velocity spread of the atoms in the atomic beam and can be described by a thermal velocity distribution with the following form:

\begin{equation}
f(\lambda) = N\left(\frac{\lambda}{\lambda_0}-1\right)^3 \exp\left (-\frac{m c^2(\frac{\lambda}{\lambda_0}-1)^2}{2k_BT}\right )
\label{equ:beam_distribution}
\end{equation}
where $N$ is a normalisation constant, $\lambda_0$ is the resonance wavelength of the transition, $m$ is the mass of the calcium atoms and $k_{B}$ is the Boltzmann constant \cite{Ramsey}. 

The longitudinal temperature of the atoms can be obtained from fitting equation (\ref{equ:beam_distribution}) to the measured spectrum. Fig. \ref{fig:oven_temp_vs_current} shows the oven temperatures for currents in the range of 1.3~A up to 1.9~A in which the temperature increases approximately linear with the current.

In order to obtain the flux of atoms, the peak ion count rate is determined from the area of the orthogonal resonance peak. It is shown as a function of oven temperature in Fig. \ref{fig:oven_flux}. The flux of calcium atoms increases with temperature according to the relationship between evaporation rate and temperature \cite{Pilling}.

\subsection{Pulsed Laser Ablation}
\label{sec:results_ablation}
\begin{figure}
\begin{center}
\resizebox{0.50\textwidth}{!}{\includegraphics{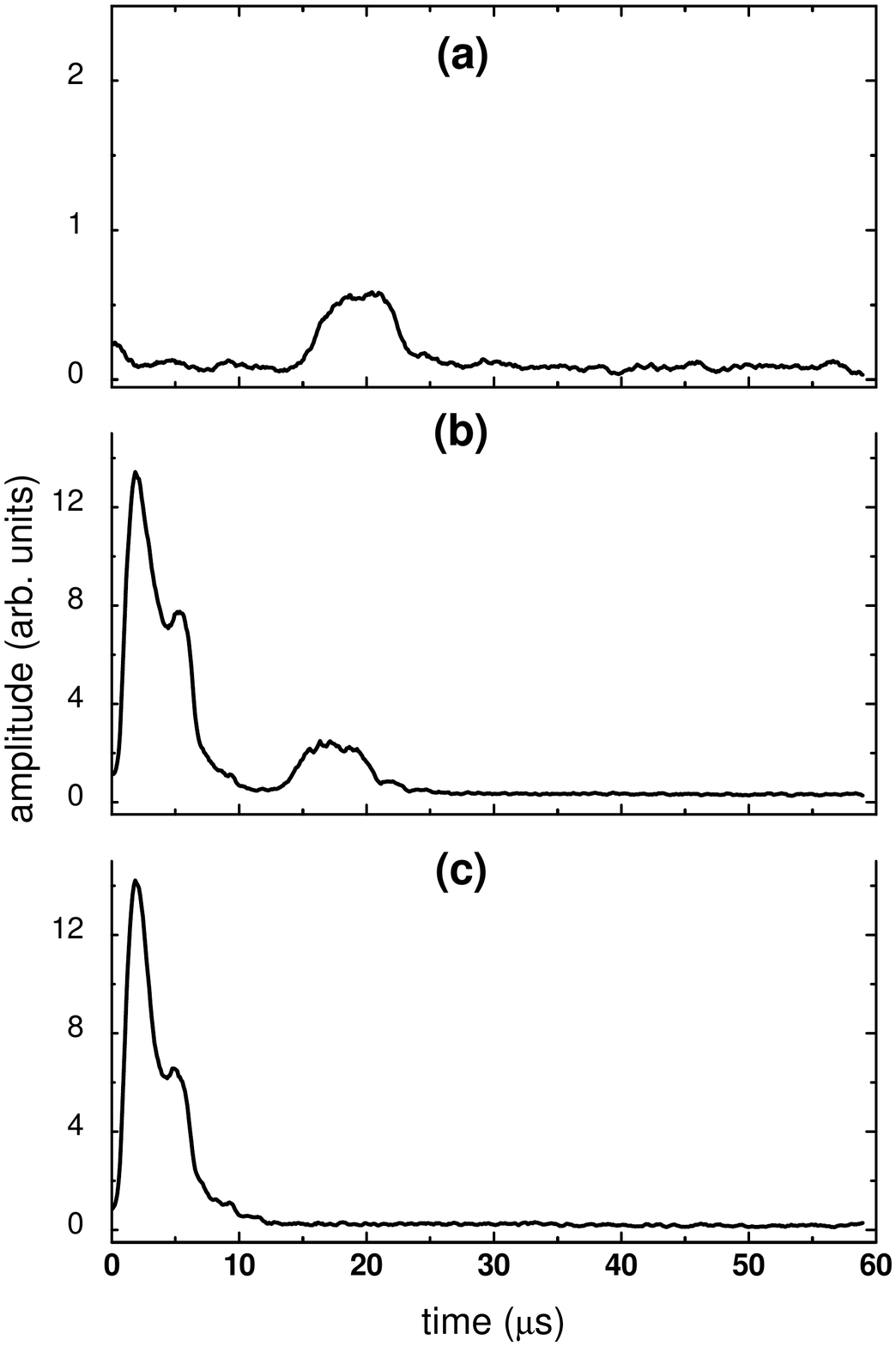}}
\end{center}
\caption{\textbf{\textit{(a)}} Spectrum of arrival times for an ablation laser fluence of 280~$\pm$~20~mJ/cm$^2$ and with the photo-ionisation lasers on. For fluences below 300~$\pm$~22~mJ/cm$^2$ the signal consists entirely of ground state atoms. \textbf{\textit{(b)}} Spectrum of arrival times recorded for an ablation laser fluence of 480~$\pm$~34~mJ/cm$^2$. The signal consists of both a ground state peak, located to right of 12~$\mu$s, and a double peak structure, located to the left of 12~$\mu$s, made up of both ions and excited state atoms. \textbf{\textit{(c)}} Spectrum collected for the same parameters as in (b) but with the resonant 423~nm laser switched off. The signal consists of entirely non-ground state atoms with times-of-flight no greater than 12~$\mu$s. }
\label{fig:ablation_time_plots}      
\end{figure}

In order to measure the velocity distribution of the atomic beam created by PLA, the time of flight of the atoms was measured.
Fig. \ref{fig:ablation_time_plots} shows time-of-flight spectra collected for two different laser fluences. \ref{fig:ablation_time_plots}(a) shows the spectrum of photo-ionised calcium atoms recorded for an ablation laser fluence of 280~$\pm$~20~mJ/cm$^2$. The atomic flux produced with this laser fluence consists entirely of ground state atoms. The composition of the atomic beam was carefully analysed by applying a voltage to the target in order to suppress the ionic fraction of the beam and by switching the resonant photo-ionisation laser (at 423~nm) to identify the neutral calcium ions in excited states. Non-ground state atoms can still be ionised through field ionisation by the capacitor field or by absorbing a 375~nm photon from the non-resonant photo-ionisation laser whereas ground state atoms won't be ionised in the absence of the resonant laser field at 423~nm. The analysis of several measurements show that atoms arriving in a time window between around 12~$\mu$s and 30~$\mu$s are all in the ground state. However, atoms arriving at earlier times are either in an excited state or are ions produced in the ablation process. Hence, the ground state atoms can be singled out by their arrival time.

As the ablation laser pulse fluence is increased, excited state atoms are created in addition to the ground state atoms. Hendricks et al. describe the creation of Rydberg atoms for a laser pulse fluence of 300~mJ/cm$^2$ \cite{Hendricks}. The results presented here confirm the creation of non-ground state atoms, both excited state atoms and ions, beginning at fluences between 300~$\pm$~22~mJ/cm$^2$ and 330~$\pm$~23~mJ/cm$^2$. Fig. \ref{fig:ablation_time_plots}(b) shows the time-of-flight spectrum collected for an ablation laser fluence of 480~$\pm$~34~mJ/cm$^2$. The double peak structure located at arrival times shorter than 12~$\mu$s consists of fast ions and slower excited state atoms which are ionised either by the electric field or through absorption of non-resonant photons at 375~nm. The distribution of ground state arrival times is visible between 12~$\mu$s and 25~$\mu$s. Fig. \ref{fig:ablation_time_plots}(c) shows the same ablation pulse as in Fig. \ref{fig:ablation_time_plots}(b) but without the resonant 423~nm laser, thus eliminating the ground state component of the total spectrum.

\begin{figure}
\begin{center}
\resizebox{0.50\textwidth}{!}{\includegraphics{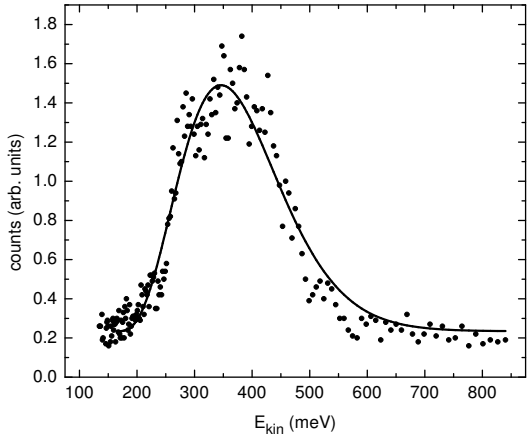}}
\end{center}
\caption{Distribution of photo-ionised calcium atoms produced by an ablation laser fluence of 480~$\pm$~34~mJ/cm$^2$. The time axis is converted to a kinetic energy axis. Eq. (\ref{equ:beam_distribution}) is fit to the distribution and gives a temperature of 2680~$\pm$~67~K corresponding to a most probable speed of 1300~$\pm$~204~m/s and a most probable kinetic energy of 340~$\pm$~77~meV.}
\label{fig:475_Maxwell}     
\end{figure}

\begin{figure}
\begin{center}
\resizebox{0.50\textwidth}{!}{\includegraphics{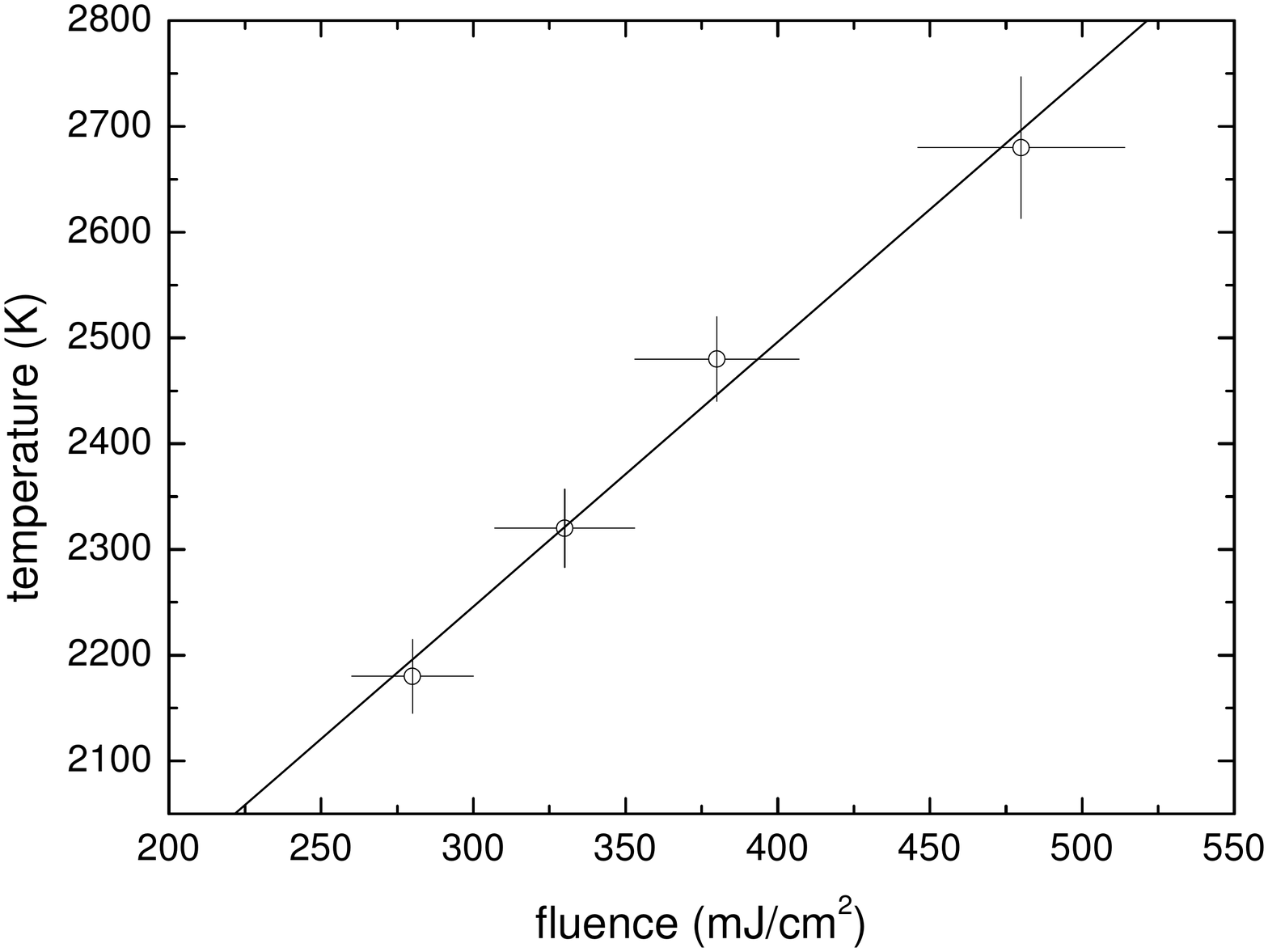}}
\end{center}
\caption{Temperature of the ablated atoms as a function of ablation laser pulse fluence. The temperature is measured by fitting a Maxwell-Boltzmann velocity distribution to the spectrum of photo-ionised atoms recorded for each ablation laser fluence. The relationship between temperature and fluence is linear for the range of fluences used in this experiment.}
\label{fig:temp_vs_fluence}     
\end{figure}

A Maxwell-Boltzmann distribution is fit to each ground state arrival time spectrum in order to determine the temperature of the atoms and their most probable kinetic energy. Fig. \ref{fig:475_Maxwell} shows the fit for a ground state time-of-flight measurement for an ablation laser fluence of 480~$\pm$~34~mJ/cm$^2$. The relationship between the temperature of the atoms and ablation laser fluence is shown in Fig. \ref{fig:temp_vs_fluence}. In the range of fluences used in this work, the beam temperature increases linearly with increasing fluence. The temperatures are typically three to four times higher than for atomic beams created by a resistively heated oven.

\begin{figure}
\begin{center}
\resizebox{0.50\textwidth}{!}{\includegraphics{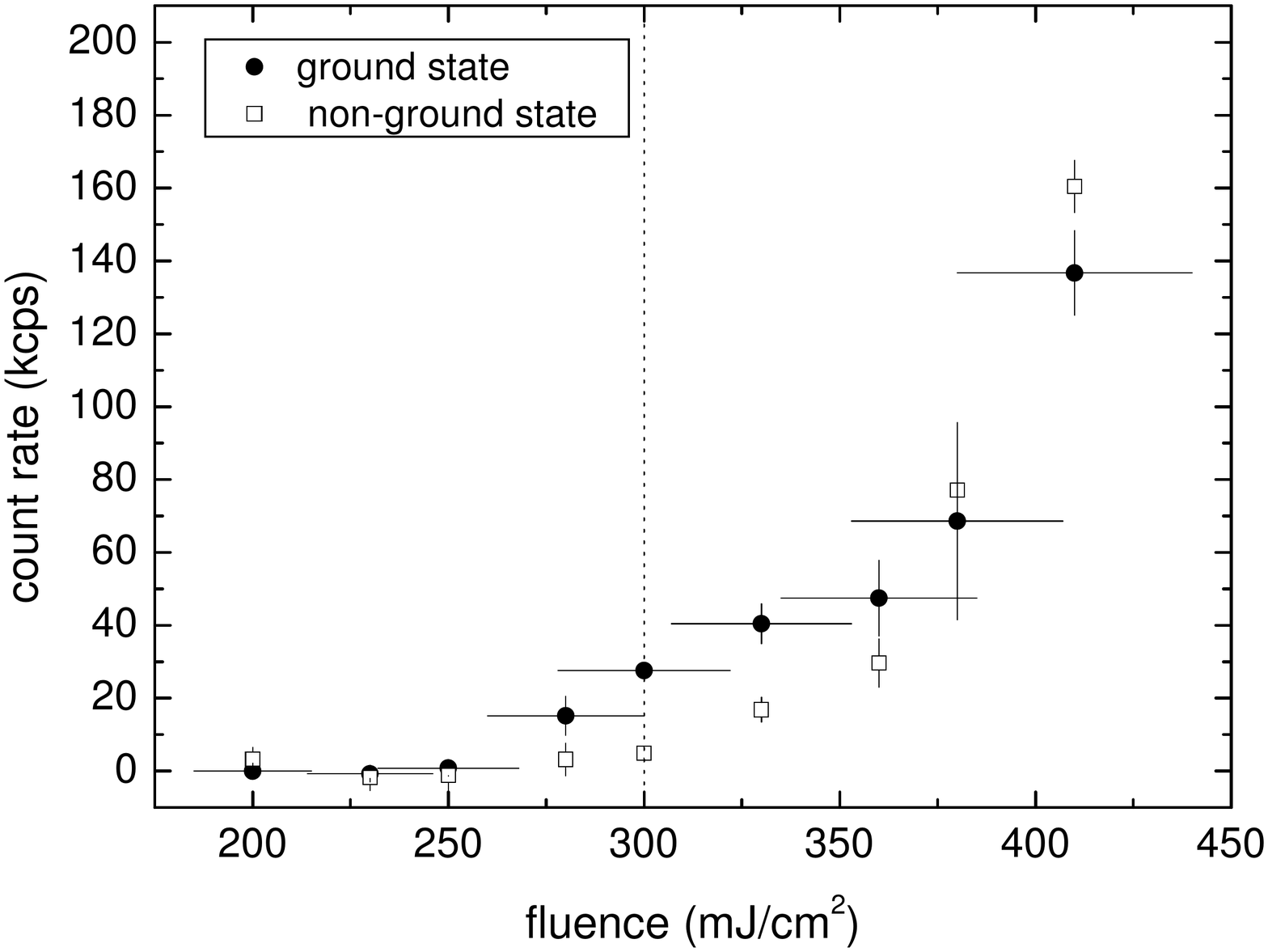}}
\end{center}
\caption{The ground state signal and non-ground state noise atomic flux production rates for fluences in the range of 210~$\pm$~15~mJ/cm$^2$ to 410~$\pm$~30~mJ/cm$^2$. For fluence below around 300~mJ/cm$^2$ the signal is comprised entirely of ground state atoms. This region, to the left of the dashed line, is the ideal region for producing atomic beams in an isotope selective ion trap loading scheme. Above this fluence the atomic beam contains both ground and non-ground state atoms and the benefits of producing atomic flux through PLA are reduced.}
\label{fig:pla_count_rates}       
\end{figure}

Separate measurements were performed to determine the relative fluxes of ground-state and non ground-state particles as a function of ablation laser fluence. From time-of-flight spectra for a range of fluences the ground-state and non ground-state fluxes are extracted from the total count rates in the time window from 12~$\mu$s to 24~$\mu$s for the ground-state flux and 0~$\mu$s to 12~$\mu$s for the non ground-state flux. The background count rate of the channeltron detector is deducted from the measurement. The ground state and non-ground state count rates as a function of ablation laser fluence are plotted in Fig. \ref{fig:pla_count_rates}. A ground state count rate above background noise is measured for fluences greater than 280~$\pm$~20~mJ/cm$^2$. For these fluences, the signal is made up entirely of ground state atoms. A non-ground state signal appears at fluences greater than 300~$\pm$~22~mJ/cm$^2$ and increases rapidly with fluence and exceeds the ground state signal for fluences larger than 360~$\pm$~25~mJ/cm$^2$. For larger fluences, ground state atoms still form a portion of the total atomic flux as is illustrated in Fig. \ref{fig:ablation_time_plots}(b). However, the dominant non-ground state fraction precludes isotope selective ion trap loading.

%%%COMPARISON%%%
%%%%%%%%%%%%%%%%%%%%%%%%%%%%%%%%%%%%%%%%%%%%%%%%%%%%%%%%%%%%%%%%%%%%%%%%%%%%%%%%
\section{Comparison of Methods}
\label{sec:comparison}
The most important figures of merit for loading ion traps are the generated flux of atoms in the ground state and their kinetic energy. In order to evaluate the flux of atoms generated by the two methods, we calculate the average flux created during a single ablation pulse and compare it with the flux effusing from the oven during an identical time span. The flux of ground state calcium atoms produced during each 10~ns ablation laser pulse can only be matched for oven temperatures exceeding 650~K. At a pulse fluence of 300~$\pm$~22~mJ/cm$^2$ the average ground state atomic flux produced is 28~$\pm$~1~kcps. From Fig. \ref{fig:oven_flux} and Fig. \ref{fig:oven_temp_vs_current}, an equivalent flux may be produced by the oven at 680~K. This temperature corresponds to a total thermal load of 3.2~W. In comparison, the thermal load delivered to the target surface by the ablation laser is 230~$\pm$~2~mW, which is more than one order of magnitude smaller. This may be a decisive advantage for use with micro-scale ion trap arrays or integrated ion chips. Table \ref{tab:equiv_oven_temp} shows the average ground state flux produced by PLA and the oven temperature required to achieve the equivalent flux. The temperatures are calculated by extrapolating the fit in Fig. \ref{fig:oven_flux}.

The resistively heated oven delivers a continuous flux in contrast to the ablation laser which creates discrete bunches of atoms. This is a considerable advantage as the pulsed ablation laser beam may be blocked following the successful loading of an ion, thus immediately halting the atomic flux. This may reduce the contamination of the trap structure significantly. In other applications, the high repetition rate for loading the trap may be a crucial advantage. Given the fact that the loading cycle with an oven is determined by slow thermal conduction, experiments seeking to load single ions with kHz cycle time may only be possible using PLA.

\begin{figure}
\begin{center}
\resizebox{0.55\textwidth}{!}{\includegraphics{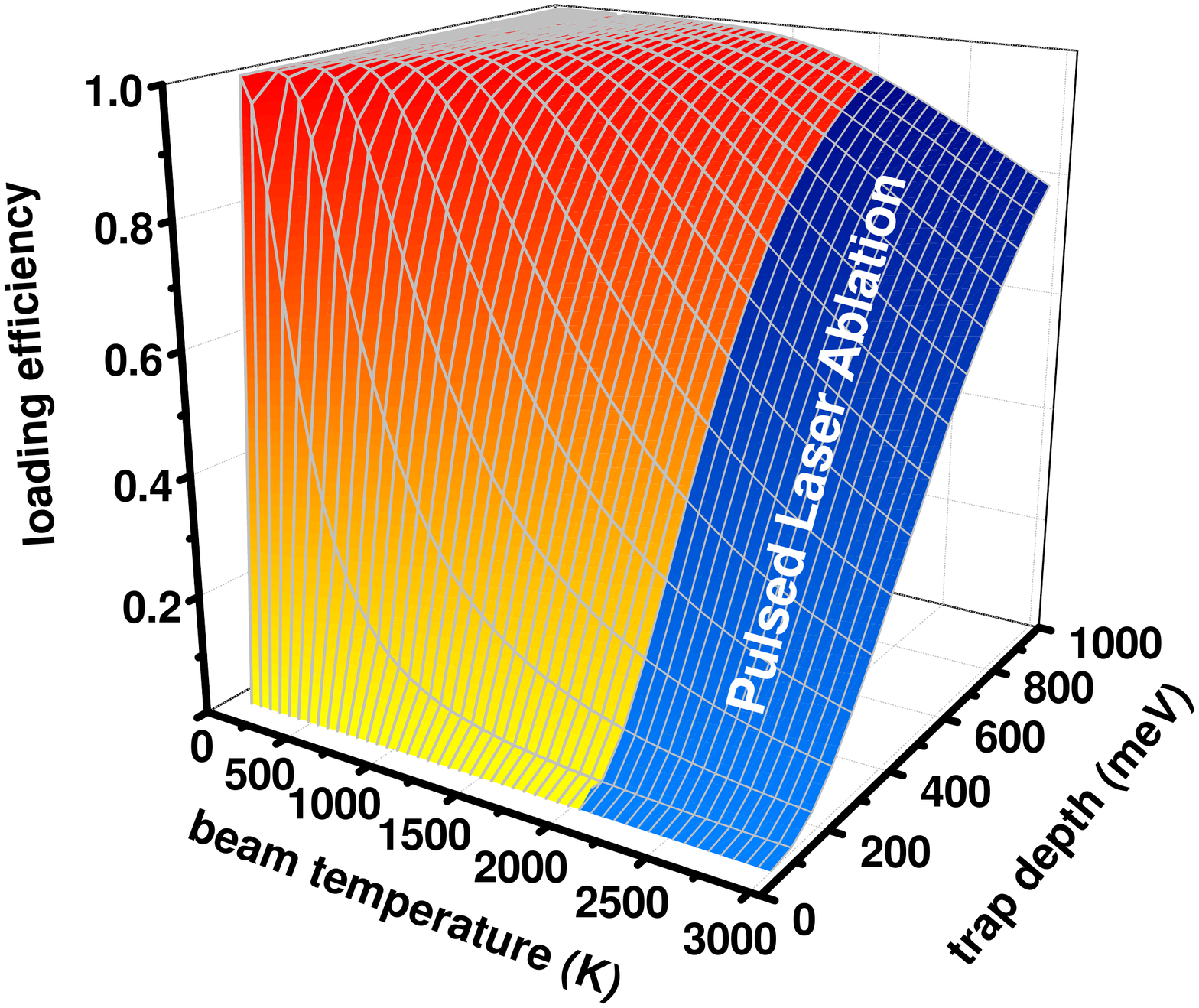}}
\end{center}
\caption{Fraction of the atomic flux which can be potentially trapped for a range of beam temperatures and trap depths. The atomic beam temperature range accessible to PLA is indicated in the diagram.}      
\label{fig:loading_efficiency} 
\end{figure}

The loading efficiency of ion traps depends on the temperature of the atomic beam as well as on the trap depth. The photoionisation process is not expected to be significantly impaired by the increased atomic velocities for intense ionization lasers orthogonally aligned with respect to the atomic beam. In our investigation, we found that the width of the atomic ionisation spectrum was the same for a conventional oven and PLA, limited only by saturation broadening of the transition. Thus, the residual Doppler broadening of the resonant transition was less than 100~MHz. The trap depth, however, limits the range of the kinetic energy spectrum of the atomic beam which can be utilized for ion trap loading. If ionised within the trap centre, only ions with a kinetic energy below the depth of the trapping potential are confined. With a significantly increased atomic beam temperature from PLA, the loading efficiency is greatly reduced for shallow traps. Figure \ref{fig:loading_efficiency} shows the fraction of atoms with a kinetic energy less than the trap depth for a range of atomic beam temperatures and trap depths. The temperature range of PLA is indicated in the graph. 
For a shallow ion trap with a depth of 100~meV only 4\% of the atoms generated by PLA can be trapped as ions. In comparison, 34\% of atoms from an effusive oven will be trapped for an atomic beam temperature of 700~K which corresponds to a comparable flux. However, already for a trap depth of 700~meV more than 80\% of the atomic flux produced by PLA can be trapped.

\begin{table}
\caption{Ground state count rates calculated for each ablation laser pulse fluence and resistively heated oven temperatures required to produce the equivalent atomic flux.}
\label{tab:equiv_oven_temp}       
\begin{center}
\begin{tabular}{clc}
\hline\noalign{\smallskip}
fluence (mJ/cm$^2$)  & count rate (kcps) & oven temp (K) \\
\noalign{\smallskip}\hline\noalign{\smallskip}
280 $\pm$ 20 & 15 $\pm$ 5 &  660 \\  \hline
300 $\pm$ 22 & 28 $\pm$ 1 & 680 \\  \hline
330 $\pm$ 23 & 40 $\pm$ 5 & 700 \\  \hline
360 $\pm$ 25 & 50 $\pm$ 10 & 710 \\  \hline
380 $\pm$ 27 & 70 $\pm$ 26 &  720 \\  \hline
410 $\pm$ 30 & 140 $\pm$ 11 & 740 \\  \hline
\noalign{\smallskip}\hline
\end{tabular}
\end{center}
\end{table}

%%%CONCLUSIONS%%%
%%%%%%%%%%%%%%%%%%%%%%%%%%%%%%%%%%%%%%%%%%%%%%%%%%%%%%%%%%%%%%%%%%%%%%%%%%%%%%%%
\section{Conclusions}

We have investigated the generation of atomic beams through PLA with 10~ns Nd:YAG laser pulses. The atoms have a Maxwellian velocity distribution with a temperature above the boiling point of calcium which increases linearly with the pulse fluence for the investigated range. The velocities are roughly a factor of 3-4 greater than those of atomic beams produced by a resistively heated oven. The corresponding kinetic energy is less than 350 meV, still well below the depth of typical macroscopic ion traps. However, for shallow traps the loading efficiency is significantly impaired.

Below a threshold fluence of 330~$\pm$~23~mJ/cm$^2$, the atomic beam produced through PLA is shown to consist of exclusively ground state atoms. Maintaining a pulse fluence below this threshold ensures that a precisely controlled flux of ground state atoms is made available for isotope selective photo-ionisation.

In conclusion, PLA in combination with resonant photoionisation appears to be a viable alternative for loading ion traps. It offers unparalleled temporal control of the loading process as the production of ground state atoms can be started or halted instantaneously. For shallow ion traps, the high atomic beam temperatures lead to a reduced loading efficiency which may offset its advantages.

%%%Acknowledgements%%%
%%%%%%%%%%%%%%%%%%%%%%%%%%%%%%%%%%%%%%%%%%%%%%%%%%%%%%%%%%%%%%%%%%%%%%%%%%%%%%%%
\section*{Acknowledgements}
This work was supported by EPSRC and the Marie Curie Actions program.

\end{document}